\documentclass[aps,pra,twocolumn,groupedaddress]{revtex4-1}
\usepackage{graphics,graphicx,float,amsmath, xcolor} 
\usepackage{upgreek}
\usepackage{hyperref}

\begin{document}

\title{Efficient signal processing for time-resolved fluorescence detection of nitrogen-vacancy spins in diamond}

\author{A. Gupta}
\affiliation{Department of Physics, McGill University, 3600 Rue University, Montreal QC H3A 2T8, Canada}
\affiliation{Department of Physics, Indian Institute of Technology Bombay, Powai, Mumbai - 400076, India}
\author{L. Hacquebard}
\affiliation{Department of Physics, McGill University, 3600 Rue University, Montreal QC H3A 2T8, Canada}
\author{L. Childress}
\affiliation{Department of Physics, McGill University, 3600 Rue University, Montreal QC H3A 2T8, Canada}






\begin{abstract}
Room-temperature fluorescence detection of the nitrogen-vacancy center electronic spin typically has low signal to noise, requiring long experiments to reveal an averaged signal. Here, we present a simple approach to analysis of time-resolved fluorescence data that permits an improvement in measurement precision through signal processing alone. Applying our technique to experimental data reveals an improvement in signal to noise equivalent to a 14\% increase in photon collection efficiency.  We further explore the dependence of the signal to noise ratio on excitation power, and analyze our results using a rate equation model. Our results provide a rubric for optimizing fluorescence spin detection, which has direct implications for improving precision of nitrogen-vacancy-based sensors. 
\end{abstract}



\maketitle

\section{Introduction}

The nitrogen-vacancy (NV) center in diamond has become a leading candidate for applications in precision sensing~\cite{Degen2014, Rondin2014, Childress2014} and quantum information science~\cite{Ladd2010}. Much of this interest hinges on the interplay between the electronic spin and the optical transitions of the defect, which permit preparation and fluorescence detection of the spin, even in ambient conditions. 
Nevertheless, the low signal to noise of the standard room-temperature fluorescence detection poses challenges for applications.  For example, in typical sensing applications, fluorescence shot noise - and not spin projection noise - limits measurement precision~\cite{Rondin2014}. There exist other approaches that can significantly improve spin measurement fidelity, such as repetitive readout~\cite{Jiang2009, Neumann2010}, resonant excitation~\cite{Robledo2011Nature}, spin-to-charge conversion~\cite{Shields2015}, and nuclear spin encoding~\cite{Steiner2011}, but they come with constraints: they increase system complexity, require longer readout times, or impose restrictions on its environment. Here, we describe an efficient and simple approach to analysis of fluorescence data that permits a small improvement in spin detection through signal processing alone, and consider the optimal regimes for its operation. 

We consider room temperature applications, in which the electronic spin detection relies on a transient optical signal. Specifically, one of the Zeeman sublevels of the $S = 1$ spin, $m_s = 0$, fluoresces more strongly than the others during the first few hundred nanoseconds of optical illumination~\cite{Doherty2013}. Current analysis techniques sum the photons observed during a fixed interval following illumination, and compare them to photon counts measured during the same interval for different prepared states. To optimize the counting interval, many experiments record time-tagged photon arrival events~\cite{Steiner2011}.  The timing information is lost when summing the photon counts. By efficiently using the time-of-arrival information, a more precise estimate of the spin populations can be made.

\section{Maximum-likelihood estimation of the spin projection}

\subsection{The mechanism for fluorescence-based spin detection}

\begin{figure}
\vspace{0in}
\includegraphics{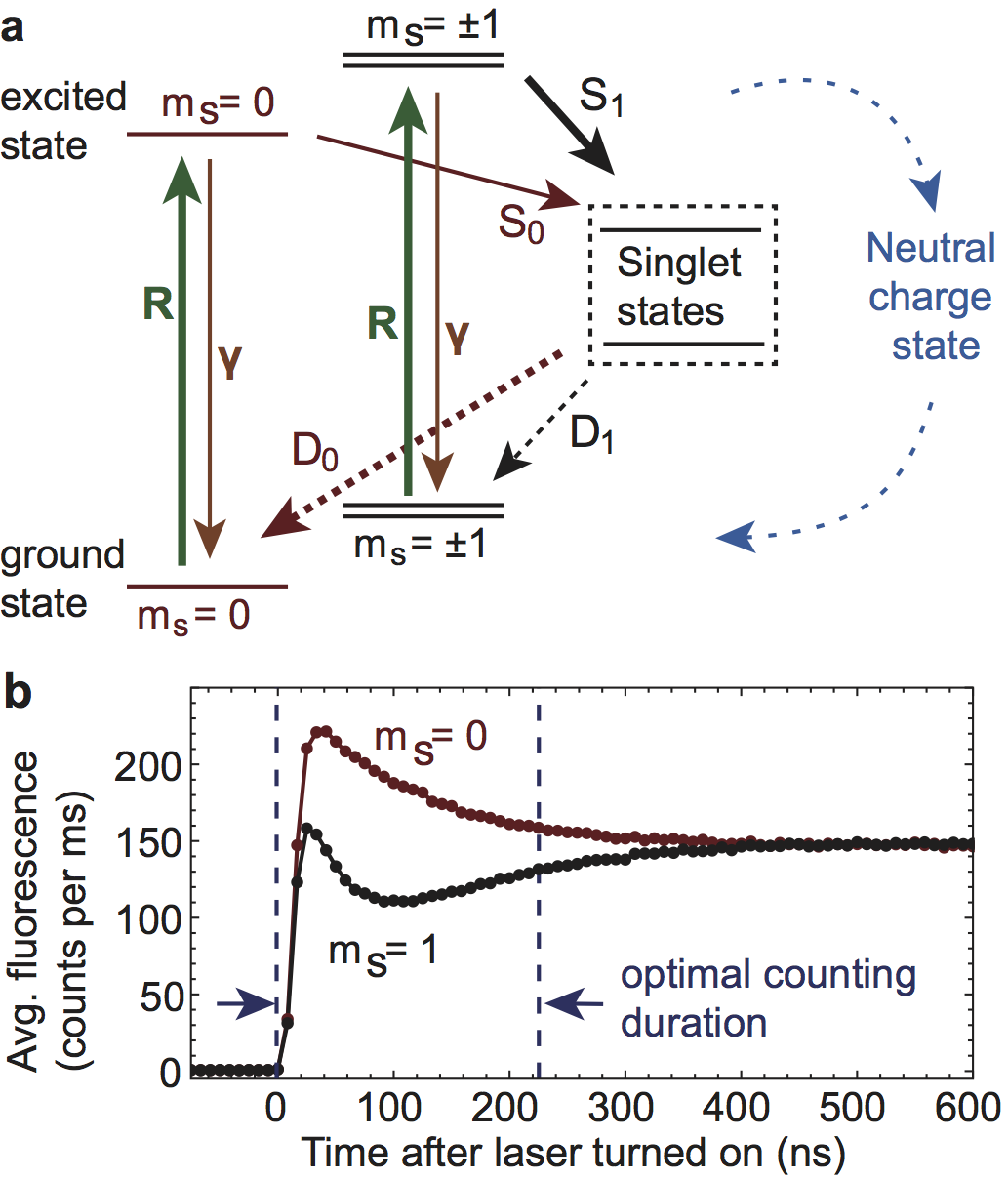}
\caption{\label{5level} (a) A simplified model for the photophysics of the NV center~\cite{Doherty2013, Robledo2011}. $R, \gamma, D_0, D_1, S_0,$ and $S_1$ are rates for  transitions indicated by adjacent arrows. In the 5-level model, transitions to the other charge state are assumed to be negligibly slow compared to other rates, and the singlet states are lumped into a single level, as are the $m_s = \pm 1$ states. (b) Example fluorescence traces after optical initialization ("$m_s = 0$") and after initialization followed by a $\pi$ pulse on the $0\rightarrow1$ spin transition ("$m_s = 1$"). Data is averaged over $N = 3*10^7$ measurement repetitions. 
In this example data, the optimal duration for the photon counting strategy is 225 ns, illustrated by the dashed lines. }
\vspace{-.2in}
\end{figure}

The basic photo-physical mechanisms behind the optical detection of the NV spin are now well established~\cite{Doherty2013}, although the details continue to be developed~\cite{Goldman2015PRL, Goldman2015}.  The essential mechanism is illustrated in the level diagram in Fig.~\ref{5level}(a). When green optical illumination is applied, the electronic transition is excited incoherently at rate $R$ via a strong phonon sideband in absorption; this process is largely spin-conserving, as is the radiative decay process, which occurs at rate $\gamma \approx 1/13$ ns~\cite{Fuchs2010, Robledo2011}. The spin-dependence of the fluorescence arises through an intersystem crossing to metastable singlet states, which occurs preferentially from the $m_s = \pm 1$ excited states. The long-lived singlet states cause the $m_s = \pm 1$ spin projection to exhibit reduced fluorescence during the first few hundred nanoseconds after optical illumination is applied. In contrast, the $m_s = 0$ spin state couples more weakly into the singlets at rate $S_0 \ll S_1$, leading to higher initial fluorescence. Furthermore, relaxation out of the singlets weakly favors the $m_s = 0$ sublevel ($D_0 > D_1$)~\cite{Robledo2011}, and the entire cycle leads to a net polarization into $m_s =  0$ of approximately $\sim70-90\%$ under continued optical illumination~\cite{Doherty2013, Fuchs2010, Robledo2011, Robledo2011Nature}.

A typical transient fluorescence signal is shown in Fig.~\ref{5level}(b), where $m_s = 0$ labels spins prepared by optical illumination and $m_s = 1$ labels spins subsequently flipped by a $\pi$ pulse. Data points are averaged over $3*10^7$ individual measurements, and it is worth emphasizing that on average a given measurement produces much less than one photon. 
Practically, the transient fluorescence signal is typically measured by counting photons in a brief period following optical illumination~\cite{Steiner2011}. As discussed below, an optimal counting time can be calculated from the time-resolved data; it is 225 ns for the example data shown in Fig.~\ref{5level}(b), as illustrated by the dashed line. Clearly, this strategy misses some of the signal, since a differential fluorescence remains after the time cutoff. Conversely, it over-weights photons that arrive towards the end of the counting interval. It is thus of interest to determine if a more efficient signal processing strategy can be employed that takes advantage of photon arrival times.

\subsection{A theoretical approach for uncorrelated photons}

To accurately measure the effect of some process on the NV center spin requires many repetitions $N$ of spin initialization, evolution, and readout. In principle, the full data set of time-resolved photon arrival times would then be required for optimal estimation of the spin projection following evolution. In most experimental situations, however, the probability is very small to detect more than one photon during a single measurement. For example, in our experiments registering $\sim200,000$ photons per second from a single NV center, on average $\sim 0.06$ photons arrive during the $\sim 300 $ ns transient spin-dependent fluorescence period, and the probability of two photons is below the percent level. Thus the majority of photons registered experimentally arrive at times that are uncorrelated with other photon arrival times (since they would belong to different measurement instances). 

To develop a simplified theory, we thus consider the limit of completely uncorrelated photons, for which the full data set can be compressed into a histogram of the number of photons arriving at some time $t$ after the readout laser turns on.  Owing to finite detector bandwidth, the histogram necessarily has a finite bin width $\Delta t$, and thus the data set can be expressed as the number of photons in each bin $\{n_1, n_2, n_3, \dots\}$. 

\subsubsection{Estimation of the spin projection}
Clearly, the fluorescence signal only permits discrimination between two of the three spin sub-levels, $m_s = 0$ vs $m_s = \pm 1$. In practice, this issue is avoided by performing experiments that isolate two of the spin sub-levels, e.g. $m_s = 0$ and $m_s = 1$. Often, one applies a magnetic field, and then employs microwave pulses that only address one of the two spin resonance transitions. For sensing applications, the microwave pulses can be used to map a quantity of interest onto the population difference $p(0) - p(1)$, where $p(0) (p(1))$ is the probability to project the state onto $m_s = 0$ ($m_s = 1$). In what follows, we thus consider only two spin sub-levels, and derive a maximum likelihood estimate for this population difference $S_z = p(0) - p(1)$. 

Suppose that we have access to a calibration that gives the mean number of photons $m_i^{(0)}$ ( $m_i^{(1)}$) that would be registered in the $i^{th}$ bin for a single readout instance performed after the spin is perfectly initialized in $m_s = 0$ ($m_s = 1$). Note that this mean number is defined for a single measurement, so it will be very small $m_i^{(0,1)} \ll 1$.  $N$ measurements are then performed on an unknown spin state with population difference $p(0) - p(1)$. After those $N$ repetitions, the total number of photons $n_i$ measured in the $i^{th}$ bin has mean value
\begin{eqnarray}
\mu_i &=& N \left(m_i^{(0)} p(0) + m_i^{(1)} p(1)\right)\\
&=&a_i + b_i S_z,
\end{eqnarray}
where $a_i = (N/2)(m_i^{(0)}  + m_i^{(1)}),$ $b_i = (N/2)(m_i^{(0)}  - m_i^{(1)}),$  and $p(0)+p(1) = 1$. Assuming Poisson processes for the photon statistics within each bin, and $m_i^{(0,1)} \ll 1$, the variance of $n_i$ is equal to its mean. Furthermore, for large $N$ the central limit theorem applies, leading to a Gaussian distribution of the number of photons in each bin. The probability that a given experimental realization $\{n_1, n_2, n_3, \dots\}$ occurs is thus 

\begin{equation}
P(\{n_i\}) = \prod\limits_i \frac{1}{\sqrt{2\pi \mu_i}}e^{- (n_i - \mu_i)^2/2 \mu_i}.
\end{equation}
 
The most likely value of $S_z$ is the value that maximizes this probability, or, equivalently, its logarithm. Differentiating $-\ln{P}$ with respect to $ S_z$  yields an equation that can be solved for the most likely value of $S_z$ given the experimental data:

\begin{equation}
\frac{d(-\ln{P})}{d S_z} = \sum_i \frac{b_i}{2}\left(1+ \frac{a_i - n_i^2 + b_i S_z}{(a_i + b_i  S_z)^2}\right) = 0.
\label{MLE}
\end{equation}

A further simplification is possible: because $b_i\ll a_i$, one might feasibly neglect the variation with $S_z$ in the denominator. Physically, this corresponds to modifying the probability distribution of each $n_i$ such that its mean $\mu_i$ varies with $S_z$ but its variance does not. Because the mean value plays a stronger role in determining the likelihood than the variance, we find that this approximation gives very good numerical results in our data analysis (see below). This approximation makes it possible to find an analytic expression for the approximate spin projection $S_z^{(A)}$: 
\begin{equation}
S_z \approx S_z^{(A)}= \frac{\sum\limits_i \frac{b_i}{a_i}\left(n_i -a_i\right)}{\sum\limits_i \frac{b_i^2}{a_i}}
\label{aMLE}
\end{equation}
It is worth noting that this approximation essentially applies a weighting function to the data according to time of arrival, with greater weight given where the signal $\sim b_i$ is larger in comparison to the noise $\sim a_i$. Since $b_i$ vanishes for large $i$, the upper limit on the sum can be any large value. 

These two estimates for $S_z$  (Eqs.~\ref{MLE}-\ref{aMLE})  can be compared to the typical approach for signal processing, in which the first $i_{max}$ bins of the data set are summed (or, equivalently, photons are counted during an interval $i_{max} * \Delta t$) to obtain an experimental signal $\eta = \sum\limits_{i = 1}^{i_{max}} n_i$. Such a photon counting approach would yield as an estimate
\begin{equation}
S_z^{(PC)} = \frac{2\left(\eta - \eta^{(1)}\right)}{\eta^{(0)}-\eta^{(1)}}-1,
\label{PC}
\end{equation}
where $\eta^{(0,1)} = \sum\limits_{i = 1}^{i_{max}} N m_i^{(0,1)} $.

\subsubsection{The signal to noise ratio}

$S_z^{(A)}$ will vary due to the noise in the Poisson-distributed $n_i$ from which it is calculated.
An expression for the variance in $S_z^{(A)}$ can thus be readily found from the probability distributions for $n_i$,
\begin{equation}
var(S_z^{(A)})  =  \sum\limits_i \left[\left(\frac{b_i}{a_i}\right)^2\left(a_i+ b_i S_z\right)\right]/\left(\sum\limits_i \frac{b_i^2}{a_i}\right)^2.
\label{varSz}
\end{equation}
If we define the signal to noise ratio (SNR) as $S_z^{(max)} - S_z^{(min)}$ (signal) divided by the square root of the variance of $S_z$ averaged over all possible values of $S_z$ (noise), then

\begin{equation}
SNR^{(A)} = 2\sqrt{\sum\limits_i \frac{b_i^2}{a_i}} = \sqrt{2N}\sqrt{\sum\limits_i\frac{\left(m_i^{(0)} - m_i^{(1)}\right)^2}{m_i^{(0)} + m_i^{(1)}}}.
\label{SNRA}
\end{equation}

This can be compared with the signal to noise ratio similarly calculated for the photon counting estimate~\cite{Steiner2011}:
\begin{equation}
SNR^{(PC)} = \sqrt{2N}\frac{\sum\limits_{i = 1}^{i_{max}}m_i^{(0)} - m_i^{(1)}}{\sqrt{\sum\limits_{i = 1}^{i_{max}}m_i^{(0)} + m_i^{(1)}}},
\label{SNRS}
\end{equation}
for which $i_{max}$ is chosen to yield the largest $SNR^{(PC)}$. Note that in all data analysis presented below, we find the optimal $i_{max}$ for each set of experimental conditions by maximizing $SNR^{(PC)}$. 

\subsubsection{Application to data}
The above theoretical treatment implicitly made several assumptions that are not always attainable experimentally. In particular, it assumed that the NV center was in its negatively charged state, in one of two possible spin states $m_s = 0$ or $m_s = 1$\footnote{$m_s = 1$ can be viewed as shorthand for $m_s = \pm1$ in zero magnetic field.}, and that one has access to calibration data $m_i^{(0)}$ and $m_i^{(1)}$ taken for perfectly initialized spin states.  For typical experiments, especially in sensing applications at room temperature, it is costly in time and equipment to initialize charge and spin with high fidelity. Fortunately, however, the same analysis holds exactly if one is interested not in the spin projection itself, but in the probability of a spin flip between the initialization and readout. 

To use the above expressions for imperfect initialization, one replaces $m_i^{(0)}$ by a calibration data set taken after initialization by optical illumination, and $m_i^{(1)}$ by a calibration data set taken after initialization plus a $\pi$ pulse on the spin transition of interest (e.g. $m_s = 0$ to $m_s = +1$). When used in a subsequent experiment (with the same initialization and readout conditions)  driving the same transition, the parameter $S_z$ calculated above is equal to $S_z = 1-2p$, where $p$ is the probability that the pulse sequence used in the experiment induced a spin flip on the calibrated transition. Essentially, this calculates the extent to which the experimental sequence mimics the $\pi$ pulse. More complex calibration techniques normally applied with the photon counting technique (see e.g. supplemental information for ~\cite{Vandersar2012}) can similarly be extended to the time resolved estimate.

\section{Experimental implementation}

By acquiring experimental fluorescence data for different prepared values of $S_z$, we can compare the three different estimators (Eqs.~\ref{MLE}, \ref{aMLE}, \ref{PC}) and examine their relative precision in analyzing the same data sets. 


We acquire data on a homebuilt confocal microscope with green (532 nm) excitation and fluorescence detection for wavelengths of 650-770 nm. The $\langle 111 \rangle$ cut chemical-vapor-deposition grown diamond sample is mounted at the focus of a NA 1.35 oil objective on an XYZ scanning piezo stage, and a 20 micron copper wire soldered across the diamond allows application of microwave signals for driving spin transitions.  The fluorescence is collected into a single mode optical fiber, and a single photon counting module ($\approx 70\%$ quantum efficiency and $<1$ ns timing resolution) converts the photons to a stream of digital pulses that encodes our signal.  

All elements of the experiment are controlled by a National Instruments 7841-R FPGA (field programmable gate array) card. The FPGA card includes sufficient analog and digital inputs and outputs to control the scanning microscope, record photon counts, and also to create digital pulse patterns to rapidly turn on and off the optical illumination and microwaves~\cite{Ziegler2012}. In addition, by overclocking the FPGA at 120 MHz, we can time-tag photon arrivals with 8.33 ns resolution, which is sufficient to resolve the transient fluorescence signals.

\begin{figure}[t]
\vspace{0in}
\includegraphics[scale = 0.6]{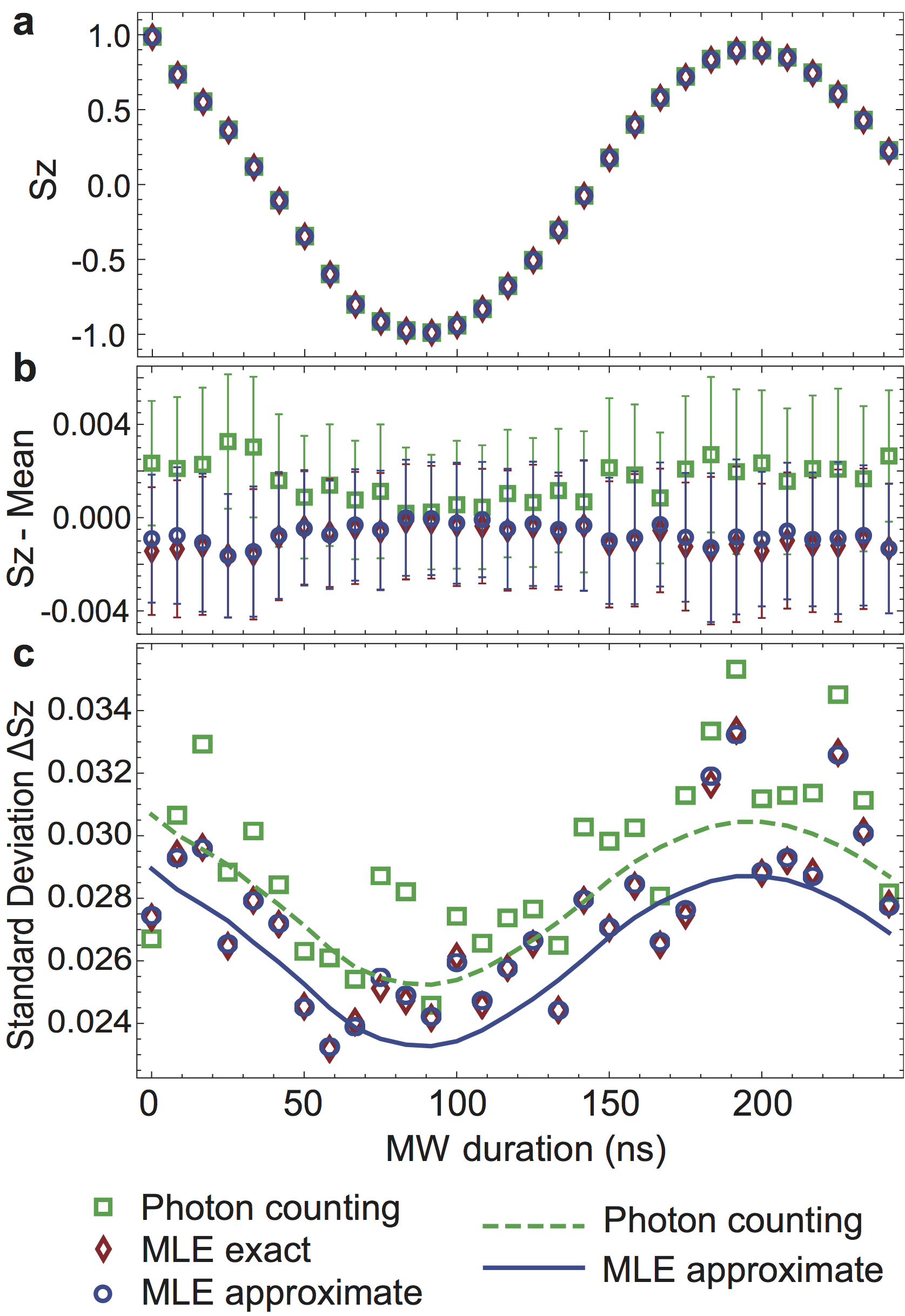}
\caption{\label{fig:SzAll} (a) The mean spin projection as estimated by the photon counting (squares), maximum likelihood (diamonds), and approximate maximum likelihood (circles) techniques over 100 repetitions of an experiment comprising $N = 10^6$ measurements. Each data point corresponds to a different duration of microwave driving following optical initialization and prior to measurement. (b) The mean value of $S_z$ (as in (a)) 
minus the average of the three methods. Error bars are standard error in the mean. (c) The standard deviation in $S_z$ (data points), along with the predicted standard deviation for the photon counting (dashed line) and approximate MLE (solid line) methods.  The legend at the bottom applies to all three parts of the figure.}
\vspace{-.2in}
\end{figure}

\subsection{Comparison of $S_z$ estimators}

We tested the data analysis techniques on three NV centers (all single defects as verified by $g^{(2)}$ measurements), and found similar results for all of them. Fig.~\ref{fig:SzAll} shows data from a representative defect (NV1) in which we analyze Rabi oscillations using the three $S_z$ estimators. A magnetic field of approximately 75 Gauss isolates the two spin transitions, and we adjust the microwave frequency and intensity such that a $\pi$ pulse on the $m_s = 0$ to $m_s = +1$ transition occurs after $91.7$ ns. We record time-tagged fluorescence data after first initializing the defect with 532 nm excitation and then driving it on this transition for variable durations, thus producing a range of possible $S_z$ values. Each microwave duration is repeated $N=10^6$ times to obtain a precise estimate of $S_z$, and then the entire experiment is performed 100 times to obtain statistics on the $S_z$ estimates.

Fig.~\ref{fig:SzAll}(a) shows the mean value of these $S_z$ estimates, on which the three techniques for analysis closely agree; the difference of each technique from the average of the three is shown in Fig.~\ref{fig:SzAll}(b), which reveals that there are deviations, but they lie within experimental error bars. In particular, the approximation for the MLE estimate is extremely close to the numeric solution to Eq.~\ref{MLE} for all data points, with the greatest deviation where $|S_z|$ is large, as expected. Fig.~\ref{fig:SzAll}(c) shows the standard deviation of the 100 $S_z$ estimates made at each microwave duration. Data points indicate the standard deviation in the values extracted by the three techniques, while the solid lines show the predicted standard deviation according to Eq.~\ref{varSz} and a similar prediction for the photon counting technique. 
The difference in noise is small, but significant. We find that the photon counting method (Eq.~\ref{PC}) has a standard deviation that is on average $7.1\pm 0.1\%$ higher than the MLE technique (Eq.~\ref{MLE}), while the approximate MLE technique (Eq.~\ref{aMLE}) is only $0.1\pm 0.1 \%$ higher.

\begin{figure}[hbt]
\vspace{.1in}
\includegraphics[scale = 0.8]{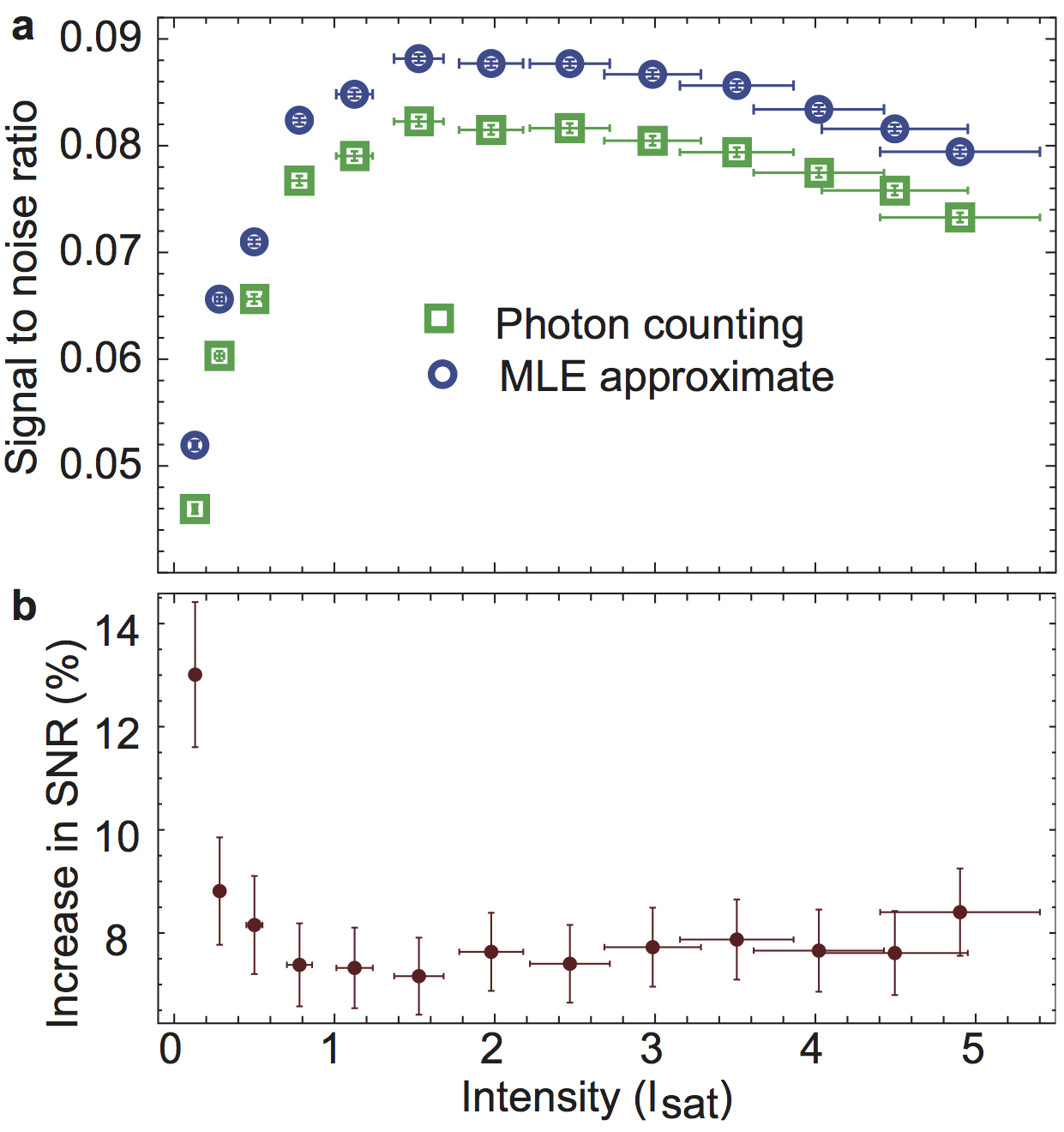}
\caption{\label{vsL}  (a) Average signal to noise ratio per run ($N = 1$) for the photon counting method and the approximate MLE method, measured for NV3 as a function of laser intensity (given in units of the saturation intensity). (b) Percent increase in signal to noise ratio associated with the MLE signal processing.}
\vspace{-.2in}
\end{figure}

\subsection{Dependence on excitation power}

The data taken in Fig.~\ref{fig:SzAll} were acquired using a readout laser intensity of approximately $2*I_{sat}$, where the saturation intensity $I_{sat}$ produces continuous-wave fluorescence rates that are half of their saturating value. In fact, the spin-dependent fluorescence signal varies with the intensity of the readout laser, as does the improvement attained with time-resolved signal processing. Fig.~\ref{vsL} shows the signal to noise ratios (Eqns \ref{SNRA} and \ref{SNRS}) and the percent difference between them as a function of laser intensity for NV3.  These data were acquired by taking fluorescence time traces for nominal $m_s = 0$ and $m_s = 1$ spin preparations at 13 different laser powers; the uncertainty in laser intensity is given as the standard deviation of values extracted from multiple saturation curves taken under nominally identical conditions. 
All three NV centers studied exhibited similar behavior, showing the highest SNR near $1.5-2*I_{sat}$, with a dramatic dropoff at lower powers and a slow decrease at higher powers. 

To quantitatively understand the SNR dependence on laser power, we used a 5-level rate-equation model~\cite{Robledo2011} in which transitions to other charge states are neglected and the singlet states are treated as a single level (see Fig.~\ref{5level}a). The parameters describing this model comprise the incoherent optical excitation rate $R$ and the radiative emission rate $\gamma$ (presumed to be equal for both spin states), as well as spin-dependent shelving rates ($S_0$, and $S_1$) and deshelving rates ($D_0$ and $D_1$), where the subscript labels the spin projection. The steady-state solution (which can be found analytically) was used to determine the initial spin polarization in the model.\footnote{More precisely, the steady-state solution was allowed to relax in the absence of optical excitation to set the initial condition of the model.}  Numerically integrating the rate equations reveals the time-dependent fluorescence rate $\propto e_0(t) + e_1(t)$, where $e_{0,1}$ are the populations in the $m_s = \{0,1\}$ optically excited states. We can then, for example, fit the numerically integrated model to our data sets to extract the model parameters. 

\begin{table*}[tb]
\begin{tabular}{cccccccccc}
 \text{NV} & \text{$t_0$ (ns)} & \text{$t_1$ (ns)} & \text{$\gamma $ (MHz)} & \text{$S_0$ (MHz)} & \text{$S_1$ (MHz)} & \text{$D_0$ (MHz)} & \text{$D_1$ (MHz)} & \text{$t_s$ (ns)} & $I_{max} (I_{sat})$ \\
 \hline
 \hline
 \text{NV1} & \text{12.94 $\pm $ 0.12} & \text{6.29 $\pm $ 0.1} & \text{67.4 $\pm $ 0.2} & \text{9.9 $\pm $ 0.8} & \text{91.6 $\pm $ 2.5} & \text{4.83 $\pm $ 0.04} & \text{2.11 $\pm $ 0.04} & \text{144. $\pm $ 1.} & $1.1\pm 0.2$ \\
 & & & \text{66.08 $\pm $ 0.09} & \text{11.2 $\pm $ 0.7} & \text{92.9 $\pm $ 2.5} & \text{4.90 $\pm $ 0.02} & \text{2.03 $\pm $ 0.01} & \text{144.3 $\pm $ 0.4} & $3.7\pm 0.7$\\
 \hline
 \text{NV2} & \text{12.93 $\pm $ 0.1} & \text{6.42 $\pm $ 0.15} & \text{67.1 $\pm $ 0.2} & \text{10.2 $\pm $ 0.6} & \text{88.6 $\pm $ 3.6} & \text{4.79 $\pm $ 0.05} & \text{2.11 $\pm $ 0.04} & \text{145. $\pm $ 1.} & $1.4 \pm 0.1$\\
& & & \text{66.43 $\pm $ 0.12} & \text{10.9 $\pm $ 0.6} & \text{89.3 $\pm $ 3.6} & \text{4.75 $\pm $ 0.03} & \text{2.13 $\pm $ 0.02} & \text{145.3 $\pm $ 0.6} & $4.6 \pm 0.4$\\
 \hline
 \text{NV3} & \text{12.95 $\pm $ 0.1} & \text{6.33 $\pm $ 0.11} & \text{65.9 $\pm $ 0.3} & \text{11.4 $\pm $ 0.7} & \text{92.1 $\pm $ 2.8} & \text{4.84 $\pm $ 0.05} & \text{2.35 $\pm $ 0.04} & \text{139. $\pm $ 1.} & $1.5\pm0.2$\\
 & & & \text{66.08 $\pm $ 0.09} & \text{11.1 $\pm $ 0.6} & \text{91.9 $\pm $ 2.7} & \text{4.90 $\pm $ 0.02} & \text{2.03 $\pm $ 0.01} & \text{144.3 $\pm $ 0.4} & $4.9\pm 0.5$\\
\hline
\end{tabular}

\caption{Parameters extracted from 5-level model fits. The first row of parameters for each defect is fit to the six lowest-intensity data traces taken for each NV center $(I_{max} < 1.5 I_{sat})$; an example is shown in Fig.~\ref{fits} for NV3; the second line is fit to all 13 data sets. The largest intensity included in the fits is noted in the final column. $t_0 = 1/(\gamma + S_0)$ and $t_1 = 1/(\gamma + S_1)$ were measured independently for each NV center, and constrained in the fits. $t_s = 1/(D_0 + D_1)$ is the singlet lifetime. All errors are standard deviation in the fit.}
\label{parameters}
\end{table*}

\begin{figure}
\vspace{.1in}
\includegraphics{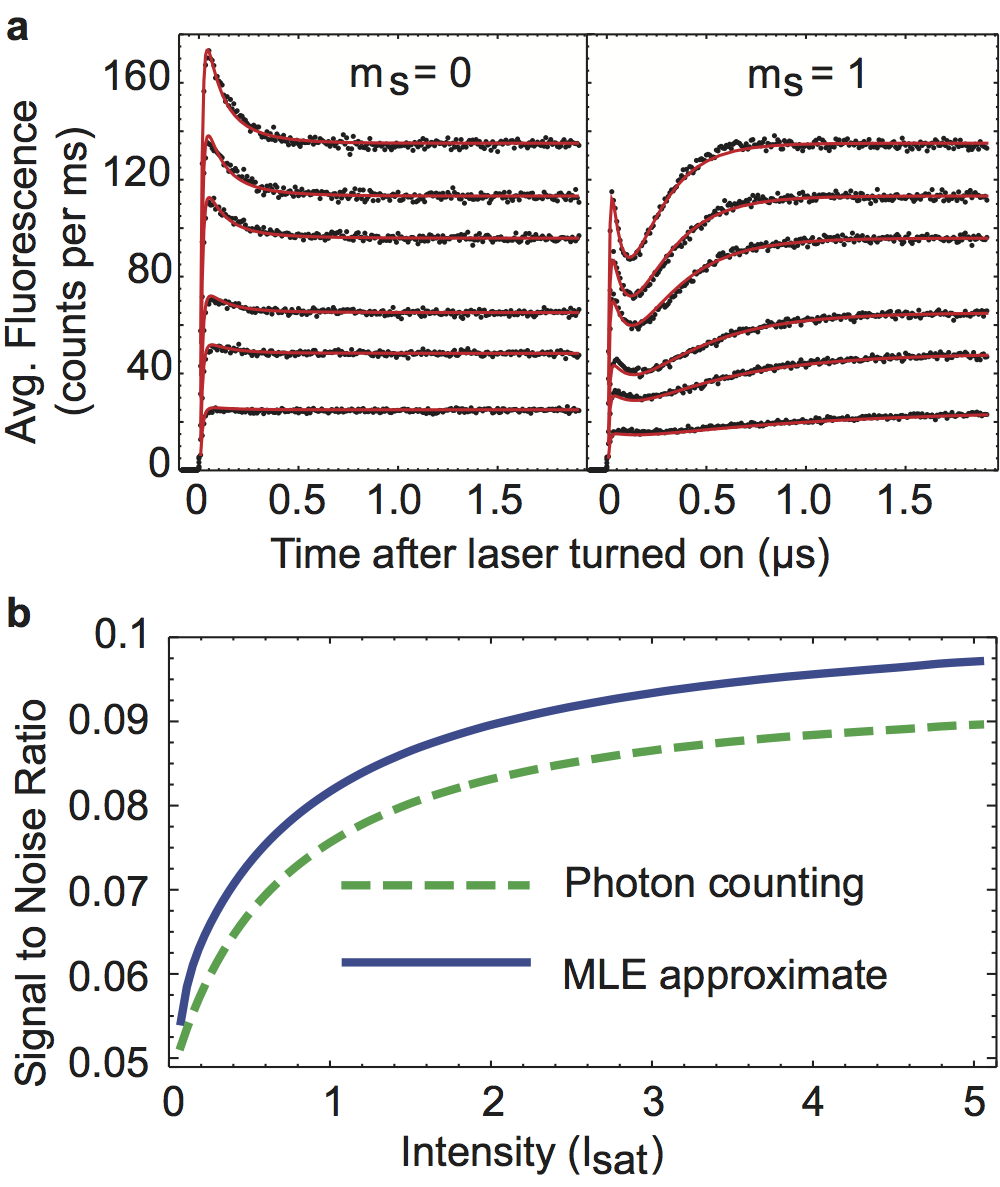}
\caption{(a) A simultaneous fit of the 5-level model to 6 data sets for NV3 acquired at different laser intensities. The lifetime of the excited states was constrained by independent measurements. For clarity, fluorescence traces are separated for nominal preparation into $m_s = 0$ (left) and $m_s = 1$ (right). (b) The signal-to-noise ratio predicted by the 5-level model using parameters extracted from the fits shown in (a).}
\vspace{-.2in}
\label{fits}
\end{figure}

In principle, only the excitation rate $R$ depends on the laser power. Thus one might expect to be able to simultaneously fit all of the data sets used in the analysis for Fig.~\ref{vsL} to the same set of parameters, only varying $R$ between the different laser intensities. We have performed such fits using a nonlinear least-squares fitting procedure including varying numbers of data sets; Fig.~\ref{fits} shows fits to data from NV3 acquired at intensities up to $1.5 I_{sat}$ (the first six data sets in Fig.~\ref{vsL}).  These fits were constrained by independent measurements of the $m_s = 0$ and $m_s = 1$ excited state lifetimes, $t_0 = 1/(\gamma + S_0) = 12.9\pm 0.1$ ns and $t_1 =1/(\gamma + S_1) = 6.3 \pm 0.1$ ns, which were extracted using a multiexponential fit to decay after pulsed excitation (data not shown)~\cite{Fuchs2010, Robledo2011}. The parameters extracted from this fit are displayed in Table~\ref{parameters}, along with similar results for other NV centers and for fits including all 13 data sets used for Fig.~\ref{vsL}. Interestingly, the resulting models predict that the optically induced polarization decreases with power, from nearly $90\%$ at low power to around $80\%$ at the higher observed intensities. However, they may not be trustworthy: we observe that some fit parameters change by more than the fit error as we include higher intensities, and the goodness of fit decreases, indicating that the 5-level model may not adequately explain our data. 

Most notably, the 5-level model cannot reproduce the observed dependence of SNR on power at high laser intensity. While initial growth with intensity is observed, the 5-level model predicts a saturating - but not decreasing - SNR at high laser power (see Fig.~\ref{fits}b). Such saturating behavior is observed for all parameter values we used in the 5-level model - even those extracted from fits to the high-power data. Some other mechanism must therefore turn on at high power to reduce signal to noise. Indeed, this is the motivation for only including low-intensity data sets (for which SNR increases with laser power) in our fits in Fig.~\ref{fits}. 

One possible explanation for the discrepancy between the 5-level model and our data is that additional parameters besides $R$ are power-dependent, perhaps through thermal effects~\cite{Toyli2012}. Another likely candidate is ionization of the NV center. Time-resolved measurements of ionization and deionization under 532 nm excitation indicate that ionization rates in the range of MHz can be expected at moderate laser powers~\cite{Waldherr2011, Chen2015}. Thus for low powers, the charge state could be modeled as approximately static during the readout, acting only to reduce the overall detected fluorescence level. At higher powers, the charge state becomes dynamic on timescales similar to the transient fluorescence signals. Dynamic charge state switching would reduce spin-dependence of the fluorescence, and could thereby account for the reduction in SNR we observe at high laser intensities.

\section{Conclusion}
By employing a simple weighting scheme derived from maximum likelihood estimation techniques, we have shown that a $7\%$ increase in signal to noise ratio for the NV spin projection can be obtained through signal processing alone. While this increase is small, we stress that many groups already employ time-resolved photon counting in order to calculate the optimal counting interval $\Delta t *i_{max}$, and thus the improvement comes for free. Furthermore, since the signal to noise ratio increases as the square root of $N$, this improvement is equivalent to what would be obtained by a $14\%$ increase in photon count rates. On the other hand, for experiments such as those using wide-field imaging onto a CCD camera~\cite{Steinert2010, Pham2011}, where photon arrival times cannot be recorded, our results are encouraging: as long as the right duration is used, the photon counting approach is very nearly as good as a maximum likelihood estimation taking into account detailed timing information. 

The formulas provided above assume uncorrelated photon arrival times, as is appropriate for standard low collection efficiency experiments. For experiments using enhanced collection efficiency techniques such as solid immersion lenses, waveguides, or other photonic devices~\cite{Hadden2010, Aharonovich2014}, the increased photon count rates will lead to a higher probability for correlations between photons. In the limit of strong photon correlations, signal processing based on quantum trajectory theory or hidden Markov models could be employed to gain further advantage in readout sensitivity and speed~\cite{Danjou2015}. Ultimately, the improvements in measurement precision gained by better signal processing translate directly to sensitivity of the NV center for magnetometry, thermometry, and other applications.








\section*{Funding Information}

This project was supported by the MITACS Globalink program, Canadian Foundation for Innovation (CFI), Canada Research Chairs, and the National Science and Engineering Research Council of Canada (NSERC). 






\bigskip

\bibliography{SignalProcessing}


\end{document}